\documentclass[12pt,preprint]{aastex}
\tighten

\def\spose#1{\hbox to 0pt{#1\hss}}
\def\n{\noindent}

\def\msun{{\rm ~M}_{\odot}}

\def\lta{\mathrel{\spose{\lower 3pt\hbox{$\mathchar"218$}}
     \raise 2.0pt\hbox{$\mathchar"13C$}}}
\def\gta{\mathrel{\spose{\lower 3pt\hbox{$\mathchar"218$}}
     \raise 2.0pt\hbox{$\mathchar"13E$}}}
\begin{document}

\title {On the Origin of X-ray Emission From Millisecond Pulsars in 47 Tuc}
\author{K. S. Cheng\altaffilmark{1} and Ronald E. Taam\altaffilmark{2}}

\affil{$^{1}$ Department of Physics,  University of Hong Kong,
Pokfulam Road, Hong Kong}

\affil{$^{2}$ Department of Physics \& Astronomy, Northwestern
University, Evanston, IL 60208}
\begin{abstract}

The observed spectra and X-ray luminosities of millisecond pulsars in 47 Tuc can be
interpreted in the context of theoretical models based on strong,
small scale multipole fields on the neutron star surface.
For multipole fields that are relatively strong as compared
to the large scale dipole field, the emitted X-rays are thermal and likely result
from polar cap heating associated with the return current from the polar gap. On the
other hand, for weak multipole fields, the emission is nonthermal and results from
synchrotron radiation of $e^{\pm}$ pairs created by curvature radiation.  The X-ray
luminosity, $L_x$, is related to the spin down power, $L_{sd}$, expressed in the
form $L_x \propto L^{\beta}_{sd}$ with $\beta \sim 0.5$ and $\sim 1$ for strong
and weak multipole fields respectively.  If the polar cap size is of the order of the
length scale of the multipole field, $s$, the polar cap temperature is $\sim 3 \times
10^6 K \left(\frac{L_{sd}}{10^{34}erg s^{-1}}\right)^{1/8} \left(\frac{s}{3\times 10^4
cm}\right)^{-1/2}$.

A comparison of the X-ray properties of millisecond pulsars in globular clusters
and in the Galactic field suggests that the emergence of relatively
strong small scale multipole fields from the neutron star interior may be
correlated with the age and evolutionary history of the underlying neutron star.
\end{abstract}


Accepted for publication in ApJ.

\section{INTRODUCTION}

The discovery of millisecond pulsars (MSPs) as a class of rapidly rotating ($P < 10$ ms),
weakly magnetized ($B \lta 10^{10}$ G) neutron stars has stimulated considerable
interest in the fundamental properties of these objects.  The detailed observational
study of these sources over periods of time have provided insights into their
origin and evolution in close binary systems (see for example, Phinney \& Kulkarni 1994).
The hypothesis that MSPs are neutron stars recycled in a spin up phase during
which angular momentum and mass are accreted from a companion star (Radhakrishnan \&
Srinivasan 1982; Alpar et al. 1982) has been dramatically confirmed with the observational
detection of the four millisecond accreting X-ray pulsars J1808.4-3658 (Wijnands \& van
Klis 1998), J1751-305 (Markwardt et al. 2002), J0929-314 (Galloway et al. 2002),
and J1807-294 (Markwardt, Smith, \& Swank 2003).  Their combination of
short spin period and low dipole magnetic field strengths have, furthermore, provided
important clues on the temporal evolution of magnetic fields
in neutron stars in low mass X-ray binary systems (van den Heuvel, van Paradijs, \&
Taam 1986; see also Bhattacharya 2002 for a recent review).

Insights into the nature of the emission mechanisms have been facilitated by
observational investigations over broad spectral regions.
As an example, the early X-ray studies of
MSPs using the ROSAT satellite revealed that the MSPs in the
Galactic field appear to have a non thermal character (see Becker \& Tr\"{u}mper 1997, 1999)
with a power law photon index ranging from $\sim -2$ to -2.4. On the other hand, the
recent X-ray studies with the Chandra satellite by Grindlay et al. (2002) indicate that the
MSPs in 47 Tuc appear to be consistent with a thermal black body spectrum characterized by
a temperature corresponding to an energy of 0.2 - 0.3 keV.

Additional evidence supporting the apparent difference between the MSPs in the Galactic
field and in 47 Tuc and, hence difference in their fundamental properties, can be gleaned
from the relation between the X-ray luminosity, $L_x$, and the spin down
power, $L_{sd}$, expressed in the form $L_x \propto L_{sd}^{\beta}$. Using ROSAT data
Becker \& Tr\"{u}mper (1997, 1999) found that $\beta \sim 1$ for MSPs in the Galactic
field, whereas there are hints that the dependence is shallower ($\beta \sim 0.5$) for
the MSPs in 47 Tuc (see Grindlay et al. 2002).  An existence of a
correlation between these two quantities provides strong evidence for relating the
energy source of the X-ray emission to the rotational energy of the underlying
neutron star.  We shall, for convenience, group the MSPs with
properties similar to the Galactic field as Type I and those similar to the MSPs in
47 Tuc as Type II, even though the nearest
MSP J0437-4715 has an X-ray spectrum consisting of two thermal components and one
non thermal component (Zavlin et al. 2002).

The conversion of rotational energy to X-ray radiation in these MSPs is likely produced
by electromagnetic processes in the neutron star's magnetosphere
(e.g., Halpern \& Ruderman 1993) rather than by frictional processes in its interior
(Alpar et al. 1984; Shibazaki \& Lamb 1988). In this case, the emission can take
place either at the magnetic poles (Daugherty \& Harding 1996) or in the outer
magnetosphere (Cheng, Ho, \& Ruderman 1986).  Specifically, it has been argued that
the non thermal X-ray emission of rotation powered pulsars results from the
synchrotron radiation of $e^{\pm}$ pairs created in the magnetosphere near the neutron
star surface by curvature photons. Such photons are emitted by charged particles on their way
from the outer magnetospheric gap to the neutron star surface (Cheng, Gil \& Zhang 1998; Cheng
\&  Zhang 1999). The non thermal X-ray luminosity is roughly about 0.1\% of the spin-down
power. We note, however, that the presence of a complicated surface magnetic field
can change the character of the emission since the
open field lines, where the outer magnetospheric gap is located, can curve
upward. In this case $e^{\pm}$ production and outflow can occur on all open field lines
and, hence, quench the outer magnetospheric gap (Ruderman \& Cheng 1988).
Observational evidence in support of
emission taking place in the magnetosphere is suggested by the existence of pulsed emission
in both the radio and soft X-ray region of the 5.75 ms pulsar J0437-4715 (Becker \&
Tr\"{u}mper 1999).

On the other hand, thermal X-ray radiation
can be produced by either neutron star cooling (Tsuruta
1998) or polar cap heating (Arons 1981; Harding, Ozernoy \& Usov 1993). Since MSPs are
extremely old pulsars, the internal heating mechanisms lead to surface temperatures
$\lta 10^5$ K (Alpar et al. 1984; Shibazaki \ Lamb 1989; Cheng et al. 1992).
Hence, the blackbody thermal emission observed from the MSPs in 47 Tuc should be attributed
to polar cap heating associated with the impact of the return current of high energy
electrons, perhaps produced in the inner or outer gaps of the magnetosphere (see Cheng, Gil, \&
Zhang 1997; Cheng \& Zhang 1999), on the neutron star surface.

However, although the X-ray emission from MSPs in the Galactic disk is dominated by non
thermal emission, the pulsed fraction, in cases that can be determined, is usually less
than 50\%. For example, the pulsed-fraction of PSR J2124-3358 is 55\% in ASCA energy
range and 33\% in ROSAT energy range respectively (Sakurai et al. 2001; Becker \&
Tr\"{u}mper 1999). Furthermore, Stappers et al. (2003) have reported an X-ray
nebula associated with PSR 1957 +20. Therefore it is possible that a significant
fraction of non thermal X-ray emission may come from an unresolved nebular component
around the pulsar. According to the observed results of Stappers et al. (2003), this
unresolved X-ray emission likely represents the shock where the winds of the pulsar
and its companion collide. Grindlay et al. (2002) have also suggested that the MSP in
NGC6397 may have such a contribution as well. On the other hand, many MSPs in 47 Tuc
have a binary companion, but their X-ray emissions are still dominated by a thermal
spectrum. Furthermore, Tennant et al. (2001) have detected X-ray emission from the
Crab pulsar at the pulse minimum. This indicates that some unpulsed-fraction can
originate from the pulsar magnetosphere. We believe that although it is possible
that the nebula may contribute to the non thermal emission, perhaps resulting in a
spectral difference between MSPs in the disk and in 47 Tuc, the observed results
have not yet provided compelling evidence to support this conjecture.

Since the spin period, binary period, X-ray luminosity, and estimated dipole magnetic field
of the two groups of MSPs overlap, other differences in properties must be sought to explain
the dichotomy.  Recently, Grindlay et al. (2002) suggested that their differences may be
related to either the existence of high order multipole fields on the neutron star
surface or to the formation of higher mass neutron stars in the dense cluster
environment of 47 Tuc. The small radius of curvature associated with high order
fields can facilitate the production of $e^{\pm}$ pair formation close to the neutron star
surface and to an increased efficiency of polar cap heating with a corresponding
increase in the level of thermal X-ray emission.  Higher mass neutron stars are more
compact and can prolong the effectiveness of the inverse Compton scattering
of thermal photons from the neutron star surface in facilitiating pair production (see
Harding, Muslimov, \& Zhang 2002) for MSPs with spin down ages $\gta 10^8$ yr.  Such a
scattering process can lead to the relation $L_x \propto  L_{sd}^{1/2}$,
however, the emission resulting from this latter process is distinctly non thermal.

We suggest, in this paper, that a small scale, strong surface magnetic field may play a
crucial role in determining the X-ray emission properties of MSPs.  The existence of such a
magnetic field may sensitively depend upon the formation history of MSPs possibly providing an
explanation for the differences between the MSPs in the field and in globular clusters. In
the next section we examine the hints provided by the observed features of MSPs.
The generic features of polar cap heating models related to
the characteristic properties of the X-ray spectrum and to the relation between its X-ray luminosity
and spin down power are presented in \S 3.  In \S 4, we compare the observed properties
of MSPs and suggest that
a possible factor differentiating these two types of MSPs is their age. The origin and
evolution of multipole magnetic fields in neutron stars is discussed within the context
of the emission models in \S 5. Finally, we summarize and discuss the implications of our
study in the concluding section.

\section{HINTS FROM OBSERVED FEATURES OF MSPS}

In the past decade, there has been significant progress in detecting and
understanding X-ray emission from rotation powered pulsars. The X-ray data
obtained from the ROSAT, ASCA, RXTE, BeppoSAX, Chandra and XMM-Newton satellite
observatories have provided very important constraints on theoretical models.
For example, Becker \& Tr\"{u}mper (1999) presented results of soft X-ray emission
from 10 MSPs in a reanalysis of archival ROSAT data, concluding that the close
correlation between the pulsar's spin-down power and the observed X-ray luminosity
suggested rotation as the energy source for the bulk of the observed non thermal
X-rays.  The linear relation between the X-ray and spin-down luminosity among
MSPs ($L_x \propto L_{sd}$) is consistent with that found in normal
radio pulsars (Becker \& Tr\"{u}mper 1997). The non thermal spectral
features of some MSPs have also been reported by Saito et al. (1997) and
Takahashi et al. (2001) based on ASCA observations and by Mineo et al. (2000)
based on BeppoSax observations. Although
the X-ray luminosity is dominated by the non thermal component, composite
spectra (power law plus black body with temperature around a few million
degrees) clearly give a better fit for the observed spectrum.  However,
the exact contribution of the thermal component to
the X-ray luminosity is difficult to determine.

Recently, Grindlay et al. (2002) presented a homogeneous data set of MSPs in 47 Tuc
observed with Chandra. This data provides a good estimate of the X-ray
luminosities and color temperatures of MSPs in 47 Tuc because these pulsars are located
at a common distance and therefore have a common interstellar column density.  This
is in contrast to the field
where the uncertainties are greater.  Although the MSPs in globular clusters share
these common quantities,
the gravitational acceleration of the globular cluster on the
MSPs contaminates the measurement of the period derivative, $\dot{P}$.
For example, about half of the MSPs in 47 Tuc
have negative $\dot{P}$ (Freire et al. 2001). While it is possible to obtain
the intrinsic $\dot{P}$ after subtraction of the gravitational effect of the
cluster by numerical modelling (see Grindlay et al. 2002)
the uncertainties in the intrinsic $\dot{P}$ of an individual MSP
can be large compared to the uncertainties in the average intrinsic value of all the MSPs
in the cluster.

Here, we show that useful information can still be gleaned from the observed
data based upon general considerations. In particular, the mean
spin-down power of MSPs in 47 Tuc can be estimated
from the typical age of these MSPs. An upper limit to their age is given by the age
the cluster, which is estimated to be $\sim$ 11 - 13 Gyr (Schiavon  et al. 2002)
based on spectroscopy and the cluster's color-magnitude diagram. A more
realistic age estimate could be derived from the age of their white dwarf
companions. Recently, Hansen, Kalogera \& Rasio (2003) suggest that the typical age of
helium white dwarfs in 47 Tuc should be $<$ 2.7 Gyr. This age gives the
mean spin-down power as $<L_{sd}> \sim \frac{I <\Omega^2>}{2}/2.7 G yr
\sim 2\times 10^{34}$ erg s$^{-1}$, where $I$ is the moment of inertia taken
to be equal to $10^{45}$ g cm$^2$ and $<\Omega>$ is the average rotational
angular frequency of the MSPs.
The mean observed X-ray luminosity of MSPs in 47 Tuc is $<L_x> =
1.95\times 10^{30}$ erg s$^{-1}$ (cf. Table 1 of Grindlay et al. 2002)  and the
ratio of these quantities is $\sim 10^{-4}$. Becker \& Tr\"{u}mper (1997,
1999) found that this ratio for normal radio pulsars as well as for MSPs, but not
including MSPs in 47 Tuc is significantly larger ($\sim 10^{-3}$).

If $L_x$ is assumed to correlate with $L^{\beta}_{sd}$, it implies
that $L_x \propto B^{2\beta}P^{-4\beta}$ where $B$ is the dipolar magnetic field
strength and $P$ is the spin period. We note that the dependence of the
X-ray luminosity is more sensitive to the spin period than to the magnetic field.
In Figure 1, $L_x$ is illustrated as a function of $1/P^2$.
The circles are MSPs in 47 Tuc, the triangles are MSPs excluding those
in 47 Tuc, and the squares are normal radio pulsars. By fitting these three sets of
data by linear regression, the slopes are $0.49\pm 0.21$, $2.16\pm0.82$ and
$1.82\pm0.45$ with correlation coefficients of 0.55(15), 0.71(9) and
0.71(18),
where the value within the parentheses corresponds to the
number of degrees of freedom. The correlation coefficients imply that the
chances of probability are $0.0335$, $0.03282$, and
$8.7\times10^{-4}$ respectively. Obviously, the data is very scattered due to the
variation of magnetic field and, hence, the chances of probability are not very
significant. However, the slopes of normal radio pulsars and MSPs
in the field are consistent with each other, whereas the slope of MSPs in 47
Tuc is clearly different.

The expected polar cap temperature of MSPs in 47 Tuc can be estimated as
$T_{exp} \sim \left(\frac{<L_x>}{\sigma_B <A_p>}\right)^{1/4} < 10^6 K$ where
$<A_p> = \pi \frac{R^3<\Omega>}{c}$ is the mean polar cap area inferred for a
dipolar magnetic field. Here, $R$ is the radius of the neutron star.  The inferred
color temperature is $\sim 3\times 10^6$ K, implying that the polar cap area is about a
hundred times
less than the expected value. The presence of a much smaller scale magnetic field
on the neutron star surface could be consistent with this result.

Finally, we re-emphasize that $L_x$ is dominated by the non thermal component
for normal radio pulsars as well as for MSPs in the field whereas $L_x$
is dominated by a black body  thermal component for MSPs in 47 Tuc. Taken as an
aggregate, these four distinguishing features suggest that the MSPs in 47 Tuc are
very distinct from their Galactic field counterparts.

In next section, we review several model predictions for the X-ray luminosities, which
must depend on the pulsar parameters, i.e. spin period and dipolar magnetic field
strength. Although the fields of the MSPs in 47 Tuc are somewhat uncertain due to the
gravitational effect of the cluster, one can roughly estimate their values.
For example, Grindlay et al. (2002) have used the King model to subtract the
gravitational effect of the cluster to obtain an estimate of the dipolar field of each
MSP in 47 Tuc. The errors in $\dot P$ are estimated to be +0.3/-0.1 in the log, which
provides the error estimates in the magnetic fields used for the calculation of
the X-ray luminosities shown in Table 1 (see \S 3).
Friere
et al. (2001) has adopted a more conservative approach and has provided upper limits of
$B$ for each MSP. In Figure 2, the observed $L_x$ versus $BP^{-2}$ is illustrated. The
circles use the field estimates of Grindlay et al (2002) and the dots use the upper
limit of Friere et al. (2001). The slopes of these two set of data are $0.92\pm 0.20$,
and $0.89\pm0.18$ with correlation coefficients of 0.78(15) and 0.81(15), which imply
that the chances of probability are $5.8\times 10^{-4}$ and $2.8\times 10^{-4}$
respectively.  Considering Figure 1 and Figure 2 together, it appears that $L_x$ is likely
proportional to $P^{-1}$ to $P^{-2}$ but the latter is more favorable because the probability of such a chance occurrence from an ensemble of systems is low.

\section{GENERIC FEATURES OF POLAR CAP HEATING MODELS}

There is a clear indication from the observed data that the spectrum of MSPs in 47 Tuc is
thermal.  However, thermal emission resulting from residual heat or frictional processes
in the interior of old neutron stars is insufficient.  The primary
mechanism for MSPs, therefore, very likely involves polar cap heating.  In this section, we review
the various polar cap heating models and compare their general features with the observed data.
Among the great number of models that have been developed to explain the pulsar radio
emission, a large fraction involve an acceleration region located near
the polar cap known as the polar gap or inner gap. Charged particles are accelerated to
relativistic energies in the polar gap, whose potential drop is limited by pair
creation. Coherent radio emission could result from the two stream instability of the
faster primary charged particle beam and the slower secondary pair beam (for a general
review, see Michel 1991). Some of these pairs created inside the polar gap can be separated
by the electric field, resulting in a backflow current.  In general,
the polar cap heating can result from this backflow current, $J_{b}$, striking the polar
caps. The X-ray luminosity is, therefore, simply given by
\begin{equation}
L_x = J_{b}V_{gap}
\end{equation}
where $V_{gap}$ is the potential difference of the polar gap.  Although the exact backflow
current to each polar cap is not known, it should be of the order of the Goldreich-Julian
current (Goldreich
\& Julian 1969), which can be written as
\begin{equation}
J_{GJ} = 1.35 \times 10^{30}B_{12}P^{-2} e s^{-1}
\end{equation}
where $e$ is the charge of electron and $B_{12}$ is the magnetic field in units of $10^{12}$
G. In other words,
\begin{equation}
J_{b} = \alpha J_{GJ}
\end{equation}
where $\alpha$ is a model dependent parameter. For a uniform pair
production situation inside the polar gap, $\alpha \sim
\frac{1}{2}$. However, this factor could be further reduced if the
current is concentrated in sparks rather than uniformly over the
polar cap (Cheng \& Ruderman 1980; Gil \& Sendyk 2000) or the
electric field in the pair creation region is actually weaker than
in other regions (Arons 1981), which is supported by statistical
analysis (e.g. Fan, Cheng \& Manchester 2001).

The predicted luminosity of the thermal X-ray radiation is rather
model-dependent since there exist a wide range of models for the polar gap potential
difference. Here, we discuss two classes of polar gap models which
depend upon whether the polar gap is sensitive or insensitive to the pulsar parameters
(viz., spin period and dipolar magnetic field). Models representative of the first group
are those described by Arons (1981) and Harding \& Muslimov (2001). Specifically, Arons (1981)
assumed the free emission of electrons (outflow) from the stellar surface with the polar
cap heating resulting from the trapped positrons (inflow) in the acceleration zone (the
polar gap) bombarding the stellar surface.  The X-ray luminosity caused by this bombardment
is estimated as
\begin{equation}
L^A_x \sim  2\times 10^{26}B_{12}P^{-27/8}f_p^{-1/4} erg ~s^{-1}
\end{equation}
where $f_p = 921 P^{1/2}s_5^{-1}$ is the ratio of the dipole radius of curvature to the
actual radius of curvature $s_5$ (in units of $10^5$ cm). Harding \& Muslimov (2001) have
included the frame dragging
effect in the emission of electron polar gap models (Scharlemann, Arons \ Fawley 1978; Arons
\& Scharlemann 1979) and predict the thermal X-ray luminosity from the polar cap as
\begin{equation}
L^{HM}_x(R) =  10^{-5} L_{sd}P^{-1/2}{\tau}_6^{3/2}
\end{equation}
if resonant inverse Compton scattering mechanism is dominant otherwise
\begin{equation}
L^{HM}_x(NR) =  10^{-5} L_{sd}P^{1/2}{\tau}_6
\end{equation}
where the spin down power of the pulsar is given by
\begin{equation}
L_{sd} = 3.8 \times 10^{31} B_{12}^2P^{-4} erg ~s^{-1},
\end{equation}
and $\tau_6$ is the characteristic age of the pulsar $\frac{P}{2\dot{P}}$ in units of
$10^6$ yr.

Another class of models predicts thermal X-ray luminosities similar to each other because
the polar gap potentials in these models are insensitive to the pulsar parameters.
For example, in the situation where ions are bound
to the polar cap surface
(Ruderman \& Sutherland 1975)
\begin{equation}
V_{RS} = 10^{12} B_{12}^{-1/7}P^{-1/7}s_6^{4/7} \rm{volts}.
\end{equation}
where $s_6 $ is the radius of curvature of the surface magnetic field in units of $10^6$ cm.
In the presence of a strong
surface magnetic field, Gil and his co-workers ( Gil \& Mitra 2001; Gil \& Melikidze 2002)
show that the
Ruderman-Sutherland potential should be modified as
\begin{equation}
V'_{RS} = \zeta^{1/7} b^{-1/7} V_{RS} \rm{volts}.
\end{equation}
where $\zeta$ is the general relativistic correction factor, which is about 0.85 for
typical neutron star parameters; $b
= B_s/B_d$, where $B_s$ is the surface magnetic field and $B_d$ is the inferred dipolar
field from the observed spin period and period derivative.

In the superstrong magnetic field B $> 0.1B_q \approx 5 \times 10^{12}$G, the high energy
photons with energy
$E_{\gamma}$ will produce electron and positron pairs at or near the kinetic threshold
(Daugherty \& Harding 1983).  Here
$E_{\gamma} = 2mc^2/sin\theta$ and $sin\theta = l_{ph}/s$ where $l_{ph}$ is the
photon mean free path for pair
formation. Cheng and Zhang (1999) argued that if the surface magnetic field is sufficiently
localized then $sin\theta =
l_{ph}/s \sim 1$ and the minimum condition for the magnetic pair production is simply
$E_{\gamma} > 2mc^2$ instead of
$ \frac{E_{\gamma}B}{2mc^2B_q} > \frac{1}{15}$ (Ruderman \& Sutherland 1975).  This
assumption yields
\begin{equation}
V_{CZ} = 1.6 \times 10^{11} s_6^{1/3} \rm{volts}.
\end{equation}

The Goldreich-Julian current will be dominated by the ion flow when the polar cap
temperature $T$ is higher than the critical temperature
\begin{equation}
T_{i} = 10^5 \eta B_{12}^{\delta} K,
\end{equation}
where the coefficients $\eta$ and $\delta$ are model dependent parameters. Jones (1986)
obtained $\eta = 0.7$ and $\delta = 0.7$ respectively, whereas Abrahams \& Shapiro (1991)
and Usov \& Melrose (1995) gave  $\eta = 3.5$ and $\delta = 0.73$ respectively. In this
case, the potential of a warm polar cap was suggested to be determined by the space charge
limited flow of ions due to the finite inertia of ions (Cheng \& Ruderman 1977; Arons \&
Scharlemann 1979). However, ions stream out from a warm neutron star surface (kT $<$ 10KeV),
and the photoejection of the most tightly bound electrons of ions (Jones 1980) acts like the
electron and positron creation mechanism to reduce the potential of the polar gap to
\begin{equation}
V_{J} = \gamma (A/Z) 10^9 volts,
\end{equation}
where $\gamma$ is the Lorentz factor of ions and A/Z is the ratio of atomic weight and
atomic number.  Typically $\gamma \sim$ 10 is required to photoeject the innermost
electrons.
It should be noted that all these potentials
(eqs.8, 9, 10 and 12)
are insensitive to the pulsar parameters. If the return current is proportional to the
Goldreich-Julian current, the functional dependence of the model luminosities on pulsar
parameters predicted by these potentials is close to $B_{12}P^{-2}$.

Cheng \& Ruderman (1980) argued that although the return current is difficult to determine,
the ion flow depends on the surface temperature exponentially. It is possible that the
return current and ion flow can adjust the surface temperature so that it is always near
the critical temperature $T_i$ shown in equation (11).
They estimated the X-ray luminosity as $L_x = \sigma_B T_i^4 A_p$ where $A_p$ is the polar
cap area. If $A_p$ is the dipolar area, the model X-ray luminosity is
\begin{equation}
L_x^{CR} =  3.7 \times 10^{24} \eta^4 B_{12}^{4\delta}P^{-1} erg ~s^{-1}.
\end{equation}

In Table 1, we compare the model predicted X-ray luminosities and the observed
data. Unless the
estimates of dipolar
magnetic fields are totally incorrect, models $L_X^{RS}$, $L_X^{CR}$, $L_X^{HM}(R)$ and
$L_X^{HM}(NR)$ are inconsistent with the observed data. The model X-ray luminosities are
higher (lower) than the observed values by more than an order of magnitude.
We have adopted $s \sim 500$ m, which is the typical
dimension of the stellar crust for realistic equations of state (Cheng \& Dai 1997)
rather than the dipolar radius of curvature.
This choice
is motivated by the work of Arons (1993), who suggested that the surface magnetic field should
be a superposition of clumps covering the entire surface of the neutron star. Further support
for such a choice is suggested by the study of Ruderman (1991a,b), who argued that the
surface magnetic field of pulsars should have a sunspot-like clump structure.  Because the core
of the neutron star is liquid, it is natural that the size of these clumps should be about
the thickness of the solid crust. In the model developed by Jones (1980), the
radius of curvature
does not enter explicitly in the potential, but implicitly through the Lorentz factor.
In this case
\begin{equation}
\gamma \approx \frac{10KeV}{3kT_{cap}(1-cos\theta_x)},
\end{equation}
where $T_{cap}$ is the polar cap temperature, which is $\sim 2.6\times 10^6$K (Grindlay
et al. 2002) and $\theta_x \sim \frac{h}{s}$ is the angle between the X-ray photon and the
local magnetic field with $h$ corresponding to the height of the polar gap. If the radius of
curvature is the dipolar value, then $\gamma \sim 10^3$, and this will not be an important
mechanism to limit the potential of the polar gap. However, assuming that $s$ is small
and hence $(1-cos\theta_x) \sim 1$, the potential is limited by the polar cap temperature.
In this approximation, we choose $\gamma \approx \frac{10KeV}{3kT} \approx
20$ and $A/Z \approx 2.$ for the model calculations.

Actually $L_x^{CR}$ and $L_x^{HM}(R)$ can be consistent with the observed values for a
much stronger dipolar
surface magnetic field ($B_{12} > 1$). However, the predicted surface temperature $T_i$ of
model $L_x^{CR}$ is still significantly lower than the observed value by an order of magnitude.
$L_x^{RS}$ can be also consistent with the observed values if the return current is
significantly lower than the Goldreich-Julian current.

In fact, all models with the predicted $L_x^{model}$ consistent with $L_X^{obs}$ require the
existence of a strong multipole field on the stellar surface including $L_x^A$. In
calculating  $L_x^A$, we have assumed the actual radius of curvature $s \sim 500m$. If
$s$ is dipolar, $L_x^A$ will increase by a factor of $\sim 10$, which makes the model
predicted luminosity higher than the observed luminosity by an order of magnitude.

With the exception of $L_x^A$, all model X-ray luminosities discussed here depend on
$1/P^{\delta}$ with $\delta$ between 1 and 2, which approximately reproduces the observed data.
However, if the magnetic field dependence is included only models $L_x^{RS}, L_x^{CZ}$ and
$L_x^{J}$ give the correct dependence on $\sim B/P^2$. These three models require small
scale, strong surface magnetic fields $> 10^{12}$G, at least near the polar cap but not
necessarily over the entire stellar
surface. This strong surface magnetic field also implies that the polar cap area is
determined by the length dimension of the surface field instead of the dipolar area. For a length
scale of the multipole field $s$, the polar cap temperature is $\sim 3\times
10^6 K \left(\frac{L_{sd}}{10^{34}erg s^{-1}}\right)^{1/8} \left(\frac{s}{3\times 10^4 cm}
\right)^{-1/2}$, which is
relatively insensitive to the spin down power and
consistent with the observed data.

\section{OBSERVATIONAL PROPERTIES OF MSPS}

As described previously, the Type I and Type II MSPs are distinct in their
X-ray characteristics, but
their overall timing properties are common.  For example, the spin periods of the
Type I X-ray sampled MSPs range from 1.56 ms - 5.26 ms (Grindlay et al. 2002) and
those of the Type II X-ray sampled MSPs range from 2.1 ms - 7.59 ms (Camilo et al.
2000). The orbital periods of those MSPs in binary systems also span a
common range as well lying between 0.38 d - 5.74 d for Type I (see Taam, King, \& Ritter 2000)
and 0.12 d - 2.36 d for Type II (Camilo et al. 2000). Although the determination of the
spin period derivative of Type II MSPs is contaminated by accelerations in the globular
cluster gravitational potential, the upper limits for the
surface dipole magnetic fields of $\lta 10^9$ G (Freire et al. 2001) are similar in
magnitude to Type I MSPs (with B in the range from $10^8 - 10^9$ G).
Thus, upon comparison of these observed and inferred properties there are no apparently
distinguishing characteristics to differentiate the X-ray properties of these
two groups.

The masses of the neutron star could be different between the two groups, but the
metallicity of 47 Tuc (with [Fe/H] = -0.7) does not significantly differ from that
of the Galactic disk to affect the properties of the neutron star (see Woosley,
Heger, \& Weaver 2002). On the other hand,
tidal capture (Bailyn \& Grindlay 1987) and exchange collisions (Rasio, Pfahl, \& Rappaport
2000) followed by a common envelope phase can provide additional channels for the formation
of binary MSPs in globular clusters.  However, the various evolutionary scenarios for
the formation of MSPs in the Galactic field (Taam et al. 2000) do not necessarily lead to
systematically different neutron star masses in the Type I group in every single case.
In fact, the fundamental issue of whether the neutron star's mass is significantly increased
during the LMXB phase is inconclusive since the amount of matter accreted can depend on
whether mass loss from the system is significant during a phase when the accretion disk
surrounding the neutron star is thermally unstable (cf., Li 2002;  Podsiadlowski, Rappaport,
\& Pfahl 2002).

Given that there are no clear distinguishing characteristics between the neutron
stars in the MSPs in the Type I and Type II groups, we consider the hypothesis that
age is a possible discriminating factor. Neutron stars in globular
clusters are likely to be older than their counterparts in the Galactic field. MSP
formation in globular clusters is likely to differ from that in the Galactic
field since the long period primordial binaries (which are the prime progenitors of short
binary period MSPs in the Galactic field) are soft in the cluster environment and, hence,
can be disrupted by stellar encounters (Heggie 1975; Hut 1984; Taam \& Lin 1992). Therefore,
the neutron stars that are present as MSPs in a globular cluster are likely to have been
isolated for as long as several Gyr, after which they underwent an exchange
collision or tidal capture to form a close interacting binary system.
The subsequent spin up evolution during an accretion
phase and the spin down evolution during the post accretion phase is not likely to be
dissimilar to those neutron stars in MSPs in the Galactic field.  Since the MSPs in the
Type I group form from primordial binaries, in contrast to those in Type II, the
neutron stars in short orbital period systems that are the primary focus of this study are
likely to have a relatively short pre accretion phase determined by the main sequence
turnoff timescale of its binary companion.  If we assume that the short period binary MSPs
are formed via the common envelope phase (for a review, see Taam \& Sandquist 2000) directly
from a progenitor system containing an intermediate mass companion (Podsiadlowski et al.
2002), then an upper limit on the duration of the pre accretion phase for MSPs of Type I can
be estimated, for example, to be $\lta 10^8$ yr for $\sim 3 \msun$ companion.

The age of the neutron star as a MSP, however, corresponds to the time since the accretion
phase ceased and can be estimated from the characteristic pulsar age given by the spin down
timescale, ${P \over 2 \dot P}$ for a braking index equal to 3. We note that this age
is only an upper limit since the spin period at the cessation of the accretion phase
may not significantly differ from its present day spin period.  For the MSPs in 47 Tuc,
the uncertainties of their characteristic ages are large and not as reliable as those
inferred for the MSPs in the Galactic field.

An observational clue for differentiating the MSPs in the two groups is provided by
the existence of 3.05 ms MSP B1821-34 in the globular cluster M28 (Lyne et al. 1987).
In contrast to the MSPs in 47 Tuc, the X-ray emission from B1821-34 is distinctly non thermal
(Becker \& Tr\"{u}mper 1997) and $L_x \propto L_{sd}$. However, its short spin
down age amounting to less than $3 \times 10^7$ years distinguishes it from the other
MSPs in the Type II group. Since its spin down timescale is much less than any of the
spin down timescales
estimated for the MSPs in 47 Tuc (see Grindlay et al. 2002), it is highly suggestive
that the neutron star age as a MSP is one factor which may help to distinguish
MSPs in Type I from those in Type II.

The 3.65 ms pulsar PSR J1740-5340 in the globular cluster NGC 6397 (D'Amico et al. 2001)
can also provide a similar constraint, however, it is unclear whether it belongs to
either of these two groups since its X-ray spectrum is non thermal, but yet it
appears that $L_x \propto L_{sd}^{1/2}$. This source is unique in that it
is an eclipsing MSP and the observations of nonvanishing emission during eclipse
(Grindlay et al. 2002) suggest that the emission region is extended.
As a result, the X-ray emission may not solely reflect processes taking place in
the immediate vicinity of the neutron star surface and the inferences drawn from
J1740-5340 are inconclusive.

The pulse timing properties MSPs in the Type I group reveal that the upper limit
of the characteristic pulsar ages is in the range from $\sim 10^9 - 3 \times 10^{10}$
yr. In principle, a better estimate of their age can be obtained from the determination
of the cooling timescales of the white dwarf companion of those MSPs in binary systems.
Amongst the MSPs in the Type I group with detectable X-ray emission, the
cooling age of the white dwarf companion in J1012+5307 has been estimated to be less than $8
\times 10^8$ yrs (Hansen \& Phinney 1998), significantly less than its pulsar characteristic
age of $>5.4 \times 10^9$ yr (Lorimer et al. 1995).  However, uncertainties
exist in such cooling ages since they sensitively depend on the thickness of the
hydrogen-rich envelope especially for the cooling of low mass helium white dwarfs.
For such white dwarfs hydrogen burning in the nondegenerate envelope can significantly
prolong its lifetime (see, for example, Sch\"{o}nberner, Driebe, \& Bl\"{o}cker 2000)
provided that thermally unstable shell flashes
do not remove the outer layers (via Roche lobe overflow) to
reduce the effectiveness of nuclear burning (Ergma, Sarna, \& Gerskevits-Antipova 2001).

Given that the available observational evidence does not discriminate between the pulsar
ages of the MSPs in Type I from those in Type II on a case by case basis, the differences
in the duration of the post accretion phase are not well constrained.  On the other hand,
it is likely that the ages of the neutron stars in 47 Tuc are comparable to the age
of the cluster corresponding to 11 - 13 Gyr (Schiavon
et al. 2002) and are older than the neutron stars in the Type I group.

\section{ORIGIN AND EVOLUTION OF MAGNETIC FIELDS}

The hypothesis of strong small scale multipole fields at the
neutron star surface provides a consistent interpretative
framework for understanding the X-ray emission properties of the
MSP of Type II.  The existence of such fields in MSP is not new,
however, since they had been suggested as possibly responsible for
the complex profiles of MSPs (Krolik 1991). Furthermore, such
fields have been hypothesized for facilitating copious production
of $e^{\pm}$ pairs required for pulsar emission in the seminal
papers of Ruderman \& Sutherland (1975) and Arons \& Scharlemann
(1979).  Thus, their presence appears to be a common component in
the phenomenology of pulsars. The creation of such fields may
arise from thermal effects occurring in the thin layers of the
crust (e.g., Blandford, Applegate, \& Hernquist 1983; Arons 1993),
from the coupling between various field components via the Hall
effect during their evolution (Shalybkov \& Urpin 1997), or from
rearrangment of the field, anchored in the core, due to crustal
movements (Ruderman 1991a, 1991b).

Although an understanding of the origin and evolution of neutron star magnetic fields remains
far from complete, observational evidence suggests that little field decay takes
place during the active lifetime of a pulsar (i.e., before the pulsar reaches the
death line) on timescales less than about a few times $10^8$ yrs (Bhattacharya et al.
1992).  Significant evolution of the magnetic field can take place, on the other hand,
during a longer pre accretion phase or during the spin up phase associated with the
accretion of matter (Konar \& Bhattacharya 1997).  Since the electrical conductivity
in the neutron star core is very high, the evolutionary timescale for the magnetic field threading
the core exceeds the Hubble timescale (Baym, Pethick, \& Pines 1969). The magnetic
field decay, therefore, reflects processes taking place in the crust. In this
paper, we adopt the hypothesis that the magnetic field decay results from ohmic
dissipation, rather than screening (Bisnovatyi-Kogan \& Komberg 1974; Taam \& van
den Heuvel 1986; Romani 1990) which is susceptible to magnetic instabilities (Bhattacharya 2002;
see also, Litwin, Brown \& Rosner 2001).
For a review on the current status of spontaneous and accretion induced
magnetic evolution in neutron stars see the recent article by Bhattacharya (2002).

Since the pre accretion phase of neutron stars in globular clusters is likely to be
significantly longer than the neutron stars in binaries in the Galactic field, it is
likely that a crustal field can decay on timescales $\sim 10^9$ yr due to the
finite electrical conductivity associated with the electron phonon scattering process
in the deep crustal layers (Konar \& Bhattacharya 1997).  In this phase, the magnetic
field can diffuse through the crust, perhaps embedding the complex field topology of
the crustal layers into the denser regions,
providing for a range of initial conditions for the field
evolution that takes place in the following accretion phase.

When the neutron star accretes matter from its binary companion (via Roche lobe
overflow) the magnetic field decay process can accelerate.  This occurs after an
isolated neutron star has acquired a companion as a result of an exchange collision
with a primordial binary or via tidal capture in a globular cluster or as a result
of the evolution of the binary itself in the Galactic field. The increase in the
interior temperatures of the neutron star resulting from the compressional heating
and nuclear burning lead to reduced electrical conductivities associated with
the electron phonon scattering process.  In contrast to the pre accretion phase,
the ion electron scattering process can play a role and be more important than
the electron phonon scattering process since the impurity content in the denser
layers from nuclear burning can be increased (Schatz et al 1999), although the
nuclear processing to Fe group nuclei that takes place during a superburst
(Schatz, Bildsten, \& Cumming 2003) may limit its importance.  As a result
a more rapid decay of the magnetic field takes place.  The decay process, however, is
limited by the depth to which matter is accreted since the electrical conductivity
increases in the denser layers which itself tends to decelerate the field decay
process.  The overall trend found by Konar \& Bhattacharya (1997) reveals
that, for a given initial dipole magnetic field, the fields decay to lower values
for lower mass accretion rates.  We note that the evolution of higher order magnetic
fields has not been calculated for the accretion phase (which may, in part, be
generated at the expense of low order fields by magnetic instabilities), although
the evolution of a
pre existing multipole field has been found to be similar to the dipole field, but on
a shorter timescale, for the
non accreting phase (Mitra, Konar, \& Bhattacharya 1999).

After the accretion phase has ceased, the recycled neutron star re-enters the pulsar
phase as a MSP with its  magnetic
field primarily residing in its core.  Any further field evolution depends on flux expulsion
from the core to the inner crustal regions as the neutron star slows down (Srinivasan et al.
1990; Bhattacharya \& Srinivasan 1995). This field may reflect its pre existing
topology, the diffusion of the field into the core during the pre accretion phase
or the displacement of the crustal field into the core accompanying the replacement of the crust
during accretion.  However, the pulsar characteristic lifetime of several
of the X-ray emitting MSPs discussed above can be long and the pulsar may not have spun
down significantly suggesting that further field decay may be minimal.  Hence, a picture
emerges whereby the field decay depends on the evolutionary history of the MSP through
its age, the amount of matter accreted by the neutron star, and the timescale on which
the accretion takes place (e.g., Bhattacharya 2002).

Based on these rudimentary models for the magnetic field evolution, we hypothesize that the
relative importance of the multipole fields is related to the long duration of the pre
accretion phase in globular clusters.  This hypothesis may be necessary, but is not
sufficient to explain the disparity between the MSPs in Type I and Type II because of the
existence of the MSPs in the globular clusters M28 and NGC 6397. If we assume that the
neutron star in the MSP in M28 is formed in the same way as those in 47 Tuc and did not
form recently as a result of an accretion induced collapse, perhaps leading to different
initial field configurations,  its existence would suggest
that the emergence of the multipole field from the core at a stage when the neutron star
is a MSP is delayed from the time that the accretion phase ceased.  In other words,
sufficient time must elapse for the multipole component of the core field to rediffuse through the
crust to the surface. Since the core cools significantly after the accretion phase and the
degree of impurities in the deep crust may be small,
the timescale for rediffusion may exceed several Gyr.

As described in the previous section, the existence of the MSP in NGC 6397 does not lend
itself to a straightforward interpretation unless it belongs to the Type I group similar
to the MSP in M28, except with a longer pulsar characteristic age.  On the other hand,
if it falls under the Type II category, then the multipole field must reemerge on
timescales of $\lta 3 - 6 \times 10^8$ yrs.  Recognizing that the pulsar ages are only
upper limits, it is possible that some neutron stars in the Type I group have a longer MSP
phase than the MSP in NGC 6397.  This would present difficulties within the above
framework and could imply that some MSP are so old that their multipole field
has decayed during the post accretion phase.  However, this may be considered unlikely
since the field decay is ultimately determined by flux expulsion from the core during
this phase.  Because the spin down timescale is so long for the very old pulsars, little
field decay is expected giving preference to the categorization for the MSP in NGC 6397
as Type I.

\section{CONCLUSIONS}

It has been shown that the X-ray properties of MSPs in 47 Tuc are distinct from
other MSPs and normal radio
pulsars.  In particular, the X-ray spectra can be described by a black body model. For
the inferred temperature and luminosity the emitting region is found to be significantly
smaller than the polar cap X-ray area deduced from a dipolar magnetic field.  In addition,
the ratio of X-ray luminosity to spin down luminosity is abnormally low (about
an order of magnitude lower), and the X-ray luminosity appears to exhibit a shallower
dependence on spin period power.
We suggest that the black body X-ray emission (kT $\sim 0.3$ keV),
from very old MSPs
(age $> 10^9$ yr) results from polar cap heating associated with the return current from the polar
gap. Such temperatures result from models where the potential difference of the polar gap is insensitive to the observed pulsar
global parameters, i.e., the rotation period and the dipolar magnetic field.  A prediction
of such models is that the thermal
X-ray luminosity of the pulsar is roughly proportional to the square root of its own
spin-down power. These models
share one similarity, namely, the existence of a very strong surface magnetic field
($> 10^{12}$ G) of very small scale ($< 10^5$ cm).
Such field strengths can follow from flux conservation arguments provided that the dipolar field lines and the surface field lines are connected, and the frame
dragging effect (Asseo \& Khechinashvili 2003) is important.
Specifically, for a polar cap area
$\sim s^2$, the surface field is $\sim 3\times10^{10}G \left(\frac{s}{3\times 10^4cm}\right)^{-2}
\left(\frac{B_d}{3\times 10^8G}\right) \left(\frac{P}{3\times 10^{-3}s}\right)^{-1}$.
Furthermore, the
general relativistic effect can amplify the surface magnetic field by a factor $>30$ for
multipole components of sufficiently high order ($\sim 5$) (see Asseo \& Khechinashvili
2003). If this interpretation is
correct it imposes a very strong hint/constraint on the evolution of MSPs.

The hypothesis of multipole fields has also been invoked as a
possible explanation for the existence of PSR J2144-3933 beyond
the pulsar death line (see Young, Manchester, \& Johnston 1999;
Gil \& Mitra 2001). In addition, Becker et al. (2003) have
recently found marginal evidence of an emission line centered at
3.3 keV from MSP PSR B1821-24. If this is identified as an
electron cyclotron line, it implies a magnetic field, at least one
hundred times stronger than its dipolar field.

We have suggested that most of the field MSPs with very old spin-down ages are probably
young MSPs (e.g., PSR B1957+20, PSR J0751+1807). Therefore, the strong multipole field may
not have had sufficient time to diffuse to the stellar surface in these pulsars. On the other
hand, some field MSPs could be very old (e.g., PSR J1012+5307, PSR J1024-0719, PSR J1744-1134),
but they may still lack a strong surface magnetic field. We speculate that this
may relate to their actual age and/or the amount mass accreted from their companions. If
the mass accreted is small, the multipole field may not anchor deeply inside the crust so
it decays on a short time scale. On the other hand, the true age of these MSPs may not be as
old as those MSPs in 47 Tuc. Confirmation of their actual age
will provide further insight into the dependence of field evolution on their
different formation histories (as hence initial magnetic field
initial conditions) and their different pre-accretion,
accretion, and post accretion phases.

\acknowledgements

\n We thank Drs. W. Becker, F. Camilo, J. Gil, W. Lewin, R.N.
Manchester, E.S. Phinney, and N. Shibazaki for helpful comments.
We are also grateful for the useful comments of the anonymous referee. This research was supported in part by a RGC grant of the Hong
Kong Government to KSC and by NASA under grant NAG5-7011 and by
the National Science Foundation under Grant No. AST-0200876 to RT.

\begin{deluxetable}{lrrrrrrrrr}
 \rotate
 \tablewidth{0pt}
 \tablecaption{Comparison between the observed X-ray luminosity and model predictions. \label{tbl-1}}
 \startdata
\cline{1-10}\\
 \colhead{} & \colhead{$L_X^{obs}$} & \multicolumn{2}{c}{$L_X^{RS}$} & \multicolumn{2}{c}{$L_X^{CZ}$}
 & \multicolumn{2}{c}{$L_X^{Jones}$} & \multicolumn{2}{c}{$L_X^{A}$} \\
 \colhead{} & \colhead{$(10^{29})$} & \multicolumn{2}{c}{$(10^{31})$} & \multicolumn{2}{c}{$(10^{29})$}
 & \multicolumn{2}{c}{$(10^{29})$} & \multicolumn{2}{c}{$(10^{29})$} \\
 \colhead{} & \colhead{erg s$^{-1}$} & \multicolumn{2}{c}{erg s$^{-1}$} & \multicolumn{2}{c}{erg s$^{-1}$}
 & \multicolumn{2}{c}{erg s$^{-1}$} & \multicolumn{2}{c}{erg s$^{-1}$} \\
 \cline{3-10}\\
 \colhead{} & \colhead{} & \colhead{(a)}   & \colhead{(b)}    & \colhead{(a)} &
 \colhead{(b)}    & \colhead{(a)}   & \colhead{(b)}    & \colhead{(a)} &
 \colhead{(b)}\\
\cline{1-10}\\
47Tuc-C & $3.98$ & $1.28 \left(^{1.72 }_{1.16 }\right)$ & $2.19 $ & $5.19 \left(^{7.33 }_{4.62 }\right)$ & $9.69 $ & $2.79 \left(^{3.95 }_{2.49 }\right)$ & $5.22 $ & $5.32 \left(^{7.52 }_{4.74 }\right)$ & $9.95 $ \\
47Tuc-D & $19.95$ & $2.16 \left(^{2.90 }_{1.96 }\right)$ & $4.62 $ & $9.22 \left(^{13.03 }_{8.22 }\right)$ & $22.39 $ & $4.97 \left(^{7.02 }_{4.43 }\right)$ & $12.06 $ & $10.54 \left(^{14.89 }_{9.40 }\right)$ & $25.59 $ \\
47Tuc-E & $39.81$ & $5.68 \left(^{7.64 }_{5.15 }\right)$ & $11.25 $ & $23.17 \left(^{32.73 }_{20.65 }\right)$ & $51.38 $ & $12.48 \left(^{17.63 }_{11.12 }\right)$ & $27.67 $ & $49.38 \left(^{69.76 }_{44.01 }\right)$ & $109.51 $ \\
47Tuc-F & $31.62$ & $5.85 \left(^{7.87 }_{5.30 }\right)$ & $19.02 $ & $20.65 \left(^{29.17 }_{18.40 }\right)$ & $81.68 $ & $11.12 \left(^{15.71 }_{9.91 }\right)$ & $43.99 $ & $68.88 \left(^{97.29 }_{61.39 }\right)$ & $272.44 $ \\
47Tuc-G & $12.59$ & $2.44 \left(^{3.28 }_{2.21 }\right)$ & $7.54 $ & $9.22 \left(^{13.03 }_{8.22 }\right)$ & $34.44 $ & $4.97 \left(^{7.02 }_{4.43 }\right)$ & $18.55 $ & $16.10 \left(^{22.74 }_{14.35 }\right)$ & $60.11 $ \\
47Tuc-H & $12.59$ & $1.64 \left(^{2.21 }_{1.49 }\right)$ & $9.25 $ & $5.19 \left(^{7.33 }_{4.62 }\right)$ & $38.96 $ & $2.79 \left(^{3.95 }_{2.49 }\right)$ & $20.99 $ & $12.78 \left(^{18.05 }_{11.39 }\right)$ & $96.02 $ \\
47Tuc-I & $15.85$ & $3.16 \left(^{4.25 }_{2.87 }\right)$ & $9.07 $ & $11.61 \left(^{16.40 }_{10.35 }\right)$ & $39.68 $ & $6.25 \left(^{8.83 }_{5.57 }\right)$ & $21.37 $ & $25.30 \left(^{35.74 }_{22.55 }\right)$ & $86.45 $ \\
47Tuc-J & $19.95$ & $6.44 \left(^{8.66 }_{5.83 }\right)$ & $10.47 $ & $20.65 \left(^{29.17 }_{18.40 }\right)$ & $36.40 $ & $11.12 \left(^{15.71 }_{9.91 }\right)$ & $19.61 $ & $96.14 \left(^{135.80 }_{85.68 }\right)$ & $169.48 $ \\
47Tuc-L & $25.12$ & $3.87 \left(^{5.20 }_{3.51 }\right)$ & $6.45 $ & $16.40 \left(^{23.17 }_{14.62 }\right)$ & $29.76 $ & $8.83 \left(^{12.48 }_{7.87 }\right)$ & $16.03 $ & $25.66 \left(^{36.25 }_{22.87 }\right)$ & $46.56 $ \\
47Tuc-M & $12.59$ & $2.08 \left(^{2.80 }_{1.89 }\right)$ & $4.46 $ & $7.33 \left(^{10.35 }_{6.53 }\right)$ & $17.82 $ & $3.95 \left(^{5.57 }_{3.52 }\right)$ & $9.60 $ & $14.73 \left(^{20.81 }_{13.13 }\right)$ & $35.84 $ \\
47Tuc-N & $15.85$ & $4.08 \left(^{5.48 }_{3.70 }\right)$ & $10.30 $ & $14.62 \left(^{20.65 }_{13.03 }\right)$ & $43.06 $ & $7.87 \left(^{11.12 }_{7.02 }\right)$ & $23.19 $ & $38.83 \left(^{54.84 }_{34.60 }\right)$ & $114.35 $ \\
47Tuc-O & $39.81$ & $5.83 \left(^{7.84 }_{5.29 }\right)$ & $18.72 $ & $20.65 \left(^{29.17 }_{18.40 }\right)$ & $80.46 $ & $11.12 \left(^{15.71 }_{9.91 }\right)$ & $43.34 $ & $68.11 \left(^{96.20 }_{60.70 }\right)$ & $265.38 $ \\
47Tuc-Q & $12.59$ & $4.41 \left(^{5.93 }_{4.00 }\right)$ & $7.57 $ & $18.40 \left(^{26.00 }_{16.40 }\right)$ & $34.56 $ & $9.91 \left(^{14.00 }_{8.83 }\right)$ & $18.62 $ & $32.21 \left(^{45.49 }_{28.70 }\right)$ & $60.48 $ \\
47Tuc-T & $10.00$ & $2.05 \left(^{2.76 }_{1.86 }\right)$ & $5.21 $ & $10.35 \left(^{14.62 }_{9.22 }\right)$ & $30.68 $ & $5.57 \left(^{7.87 }_{4.97 }\right)$ & $16.53 $ & $7.02 \left(^{9.91 }_{6.25 }\right)$ & $20.81 $ \\
47Tuc-U & $19.95$ & $3.87 \left(^{5.20 }_{3.51 }\right)$ & $8.01 $ & $16.40 \left(^{23.17 }_{14.62 }\right)$ & $38.33 $ & $8.83 \left(^{12.48 }_{7.87 }\right)$ & $20.64 $ & $25.69 \left(^{36.29 }_{22.90 }\right)$ & $60.03 $ \\
\cline{1-10}\\
\multicolumn{10}{c}{Note: (a) Grindlay et.~al.~2002 (b) Freire et.~al.~2001}\\
\cline{1-10}\\ \\ \\
\cline{1-10}\\
 \colhead{} & \colhead{$L_X^{obs}$} & \multicolumn{2}{c}{$L_X^{CR}$} & \multicolumn{2}{c}{$L_X^{HM(R)}$}
 & \multicolumn{2}{c}{$L_X^{HM(NR)}$}  & \multicolumn{2}{c}{}\\
 \colhead{} & \colhead{$(10^{29})$} & \multicolumn{2}{c}{$(10^{18})$} & \multicolumn{2}{c}{$(10^{34})$}
 & \multicolumn{2}{c}{$(10^{31})$}  & \multicolumn{2}{c}{} \\
 \colhead{} & \colhead{erg s$^{-1}$} & \multicolumn{2}{c}{erg s$^{-1}$} & \multicolumn{2}{c}{erg s$^{-1}$}
 & \multicolumn{2}{c}{erg s$^{-1}$}  & \multicolumn{2}{c}{}\\
 \cline{3-10}\\
 \colhead{} & \colhead{} & \colhead{(a)}   & \colhead{(b)}    & \colhead{(a)} &
 \colhead{(b)}    & \colhead{(a)} & \colhead{(b)} & \colhead{} & \colhead{}\\
\cline{1-10}\\
47Tuc-C& $3.98$ & $1.87 \left(^{5.14 }_{1.34 }\right)$ & $11.64 $ & $26.74 \left(^{18.93 }_{30.01 }\right)$ & $14.31 $ & $1.44 \left(^{1.44 }_{1.44 }\right)$ & $1.44 $ \\
47Tuc-D& $19.95$ & $7.11 \left(^{19.49 }_{5.08 }\right)$ & $94.68 $ & $19.34 \left(^{13.69 }_{21.70 }\right)$ & $7.97 $ & $1.61 \left(^{1.61 }_{1.61 }\right)$ & $1.61 $ \\
47Tuc-E& $39.81$ & $14.02 \left(^{38.43 }_{10.02 }\right)$ & $143.44 $ & $32.95 \left(^{23.33 }_{36.97 }\right)$ & $14.86 $ & $3.00 \left(^{3.00 }_{3.00 }\right)$ & $3.00 $ \\
47Tuc-F& $31.62$ & $2.36 \left(^{6.47 }_{1.69 }\right)$ & $130.91 $ & $105.12 \left(^{74.42 }_{117.95 }\right)$ & $26.58 $ & $4.69 \left(^{4.69 }_{4.69 }\right)$ & $4.69 $ \\
47Tuc-G& $12.59$ & $1.81 \left(^{4.97 }_{1.30 }\right)$ & $85.01 $ & $51.92 \left(^{36.76 }_{58.26 }\right)$ & $13.91 $ & $2.45 \left(^{2.45 }_{2.45 }\right)$ & $2.45 $ \\
47Tuc-H& $12.59$ & $0.11 \left(^{0.30 }_{0.08 }\right)$ & $40.05 $ & $206.49 \left(^{146.19 }_{231.69 }\right)$ & $27.49 $ & $3.46 \left(^{3.46 }_{3.46 }\right)$ & $3.46 $ \\
47Tuc-I& $15.85$ & $1.74 \left(^{4.76 }_{1.24 }\right)$ & $62.83 $ & $69.20 \left(^{48.99 }_{77.65 }\right)$ & $20.25 $ & $3.06 \left(^{3.06 }_{3.06 }\right)$ & $3.06 $ \\
47Tuc-J& $19.95$ & $0.81 \left(^{2.21 }_{0.58 }\right)$ & $4.22 $ & $228.88 \left(^{162.03 }_{256.81 }\right)$ & $129.83 $ & $6.54 \left(^{6.54 }_{6.54 }\right)$ & $6.54 $ \\
47Tuc-L& $25.12$ & $13.87 \left(^{38.03 }_{9.91 }\right)$ & $79.03 $ & $22.62 \left(^{16.01 }_{25.38 }\right)$ & $12.46 $ & $2.20 \left(^{2.20 }_{2.20 }\right)$ & $2.20 $ \\
47Tuc-M& $12.59$ & $0.59 \left(^{1.61 }_{0.42 }\right)$ & $7.87 $ & $90.94 \left(^{64.38 }_{102.04 }\right)$ & $37.38 $ & $2.83 \left(^{2.83 }_{2.83 }\right)$ & $2.83 $ \\
47Tuc-N& $15.85$ & $1.80 \left(^{4.93 }_{1.28 }\right)$ & $42.10 $ & $87.26 \left(^{61.78 }_{97.91 }\right)$ & $29.63 $ & $3.73 \left(^{3.73 }_{3.73 }\right)$ & $3.73 $ \\
47Tuc-O& $39.81$ & $2.45 \left(^{6.71 }_{1.75 }\right)$ & $129.93 $ & $102.40 \left(^{72.49 }_{114.89 }\right)$ & $26.28 $ & $4.64 \left(^{4.64 }_{4.64 }\right)$ & $4.64 $ \\
47Tuc-Q& $12.59$ & $13.52 \left(^{37.07 }_{9.66 }\right)$ & $85.16 $ & $26.19 \left(^{18.54 }_{29.38 }\right)$ & $13.94 $ & $2.46 \left(^{2.46 }_{2.46 }\right)$ & $2.46 $ \\
47Tuc-T& $10.00$ & $53.66 \left(^{147.11 }_{38.34 }\right)$ & $1282.06 $ & $5.10 \left(^{3.61 }_{5.72 }\right)$ & $1.72 $ & $0.95 \left(^{0.95 }_{0.95 }\right)$ & $0.95 $ \\
47Tuc-U& $19.95$ & $13.82 \left(^{37.89 }_{9.87 }\right)$ & $164.74 $ & $22.68 \left(^{16.06 }_{25.45 }\right)$ & $9.71 $ & $2.20 \left(^{2.20 }_{2.20 }\right)$ & $2.20 $ \\
\cline{1-10}\\
\multicolumn{10}{c}{Note: (a) Grindlay et.~al.~2002 (b) Freire et.~al.~2001}\\
 \enddata
\end{deluxetable}

\begin{figure}
\plotone{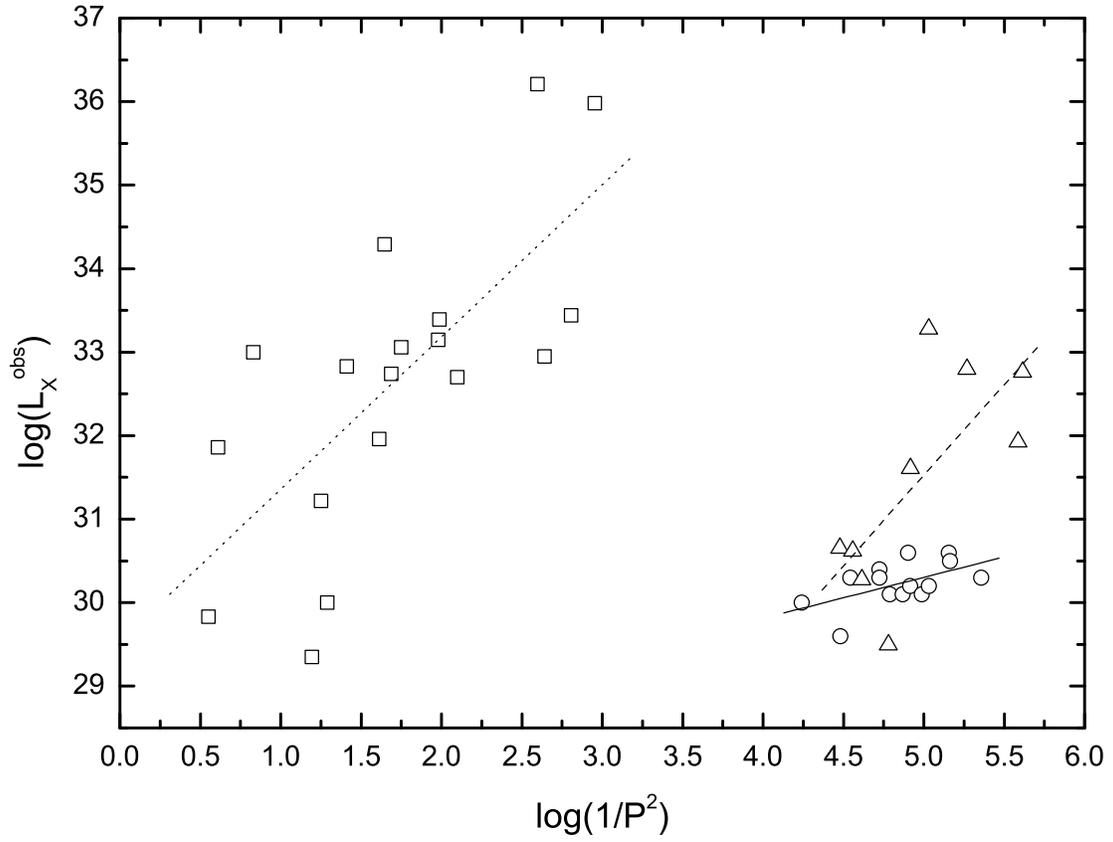}
\caption{The plot of the observed X-ray luminosity versus $1/P^2$. The squares represent
normal radio pulsars (Becker \& Tr\"umper~1997), the triangles represent MSPs excluding
those in 47 Tuc (Becker \& Tr\"umper 1999), and the circles represent MSPs in 47 Tuc
Grindlay et.~al.~2002). \label{FIG1}}
\end{figure}

\begin{figure}
\plotone{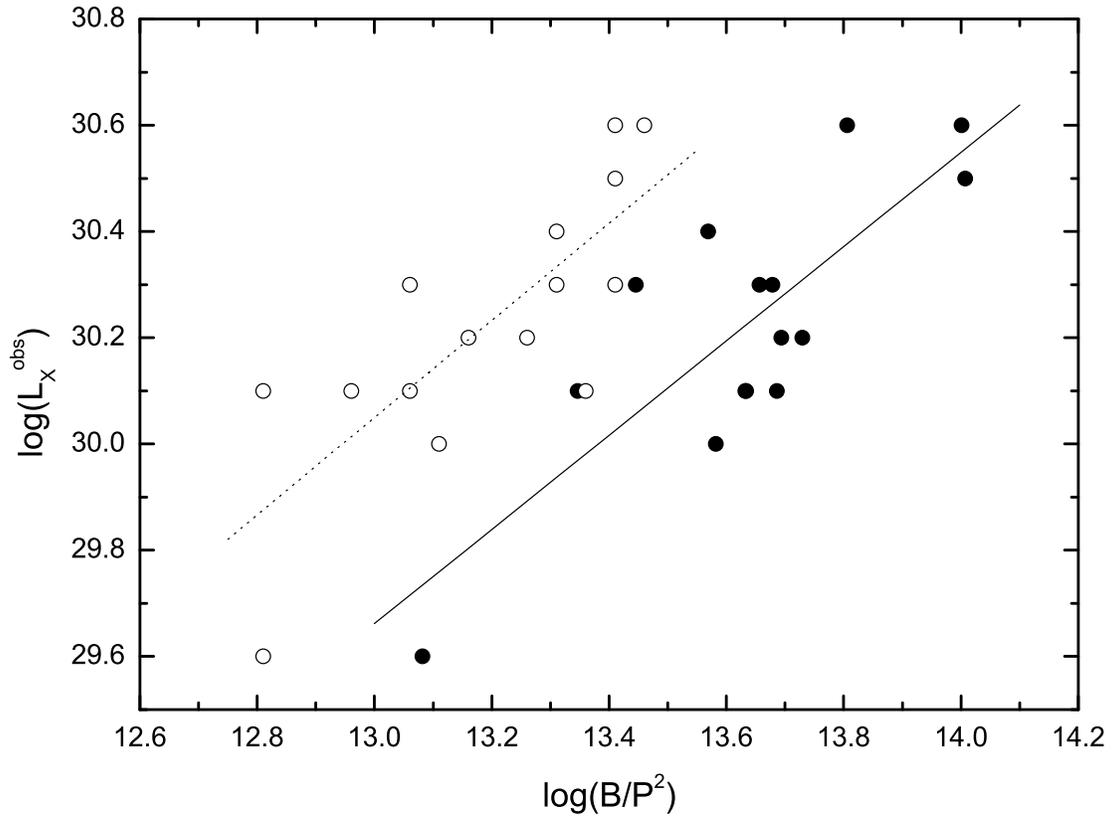}
\caption{The plot of observed X-ray luminosity versus $B/P^2$. The circles represent
data using magnetic fields estimated by Grindlay et.~al.~(2002) and dots represent data
using magnetic fields estimated by Freire et.~al.~(2001). \label{FIG2}}
\end{figure}

\end{document}